\documentclass[a4paper,twocolumn,accepted=2023-07-20]{quantumarticle}
\pdfoutput=1
\usepackage[utf8]{inputenc}
\usepackage[english]{babel}
\usepackage[T1]{fontenc}
\usepackage{amsmath}
\usepackage{hyperref}
\usepackage{amssymb}
\usepackage{amsfonts}
\newcommand{\vecg}{\boldsymbol}
\renewcommand{\vec}{\textbf}
\newcommand{\ket}[1]{|#1\rangle}
\newcommand{\bra}[1]{\langle#1|}
\newcommand{\bracket}[2]{\langle#1|#2\rangle}

\newcommand{\id}{{\mathbb{I}}}
\DeclareMathOperator{\tr}{Tr} 
\usepackage{subfig}
\usepackage{graphicx}
\usepackage{color}

\begin{document}

\bibliographystyle{quantum}

\title{Bell-type inequalities for systems of relativistic vector bosons}

\author{Alan J. Barr}
\affiliation{Department of Physics, Keble Road, University of Oxford, OX1 3RH and\\
	Merton College, Merton Street, Oxford, OX1 4JD}
\author{Pawe{\l}{} Caban}
\email{Pawel.Caban@uni.lodz.pl (corresponding author)}
\orcid{0000-0002-6973-2301}
\author{Jakub Rembieli\'nski}
\orcid{0000-0002-5379-4487}
\affiliation{Department of Theoretical Physics,
University of {\L}{\'o}d{\'z}\\
Pomorska 149/153, PL-90-236 {\L}{\'o}d{\'z}, Poland}

\maketitle


\begin{abstract}
We perform a detailed analysis of the possible violation of various Bell-type inequalities for systems of vector boson-antiboson pairs.  
Considering the general case of an overall scalar state of the bipartite system, we identify two distinct classes of such states, 
and determine the joint probabilities of spin measurement outcomes for each them. 
We calculate the expectation values of the CHSH, Mermin and CGLMP inequalities
and find that while the generalised CHSH inequality is not expected to be violated for any of the scalar states,
in the case of the Mermin and CGLMP inequalities the situation is different -- 
these inequalities can be violated in certain scalar states
while they cannot be violated in others. Moreover, the degree of violation depends on 
the relative speed of the two particles.
\end{abstract}
\maketitle

\section{Introduction}

Quantum mechanics predicts that the results of measurements exhibit correlations
that differ radically from those of classical physics. 
Einstein, Podolsky and Rosen were sufficiently disturbed by the apparent lack of realism
in quantum measurements, in particular those corresponding to non-commuting operators,
that they doubted the completeness of the theory~\cite{cab_EPR1935}. 
In response, Bell~\cite{cab_Bell1964} considered the predictions of theories that 
are both local (such the physical influences cannot travel faster than the speed of light),
and realistic (having physical properties that are independent of observation).
He showed that, under certain assumptions, one could perform experimental tests that 
could distinguish between the predictions of quantum theory and those of such local 
realistic theories. His method was based on the observation that quantum mechanical
predictions for particular correlated expectation values disobey a mathematical 
inequality  -- a so-called `Bell Inequality' -- that local realistic theories must satisfy. 

Subsequent experimental tests of Bell-like inequalities have been performed a variety 
of physics systems using e.g. photons~\cite{FC1972,ADR1982},
ions~\cite{cab_RKMSIMW2001}, 
superconducting systems~\cite{AWetal2009_josephson} and solid-state
systems~\cite{PTetal2013}. 
More recent examples include so-called `loop-hole-free' tests~\cite{HBDetal2015,GVetal2015,SMetal2015},
in which special attention was paid to issues of causality, efficiency and freedom of choice
of experimental settings.
Bell tests have also been performed experimentally on pairs photons with orbital angular
momentum~\cite{VWZ2002}, where any measurement of each of 
the two subsystems results in one of three possible outcomes.

Violation of Bell inequalities by quantum-mechanical probabilities is usually
discussed within the framework of non-relativistic quantum mechanics.
However, the description of EPR experiments with relativistic particles should be performed
in a relativistic setting. Quantum entanglement and Bell inequalities in such a setting have
been considered in the literature starting from Czachor's paper \cite{cab_Czachor1997_1}.
Since then a large number of papers on the subject have been published, see, e.g., \cite{cab_AM2002,cab_GA2002,cab_PST2002,cab_ALMH2003,cab_LD2003,cab_TU2003_1,cab_CR2005,CR2006,CRW_2008_vector_bosons,FBHH2010_Relativistic-entanglement, AF_2012_Oberver-dependent-entang,SV2012_PhysRevA.85.062101,TA_PhysRevA.88.060302,CRRSW2014_correl-localization,PD2015_Werner-relativistic,BBB2018_PhysRevA.97.032106,SGB2021_PhysRevLett.126.230403,OH2021-single-particle-entangl} 
and references therein.
It is worth taking note that the description of EPR experiments in a relativistic
framework is hindered by theoretical and interpretational difficulties. One of the most
important problems is related to the appropriate definition of a relativistic spin operator which
is rooted in the non-existence of the Lorentz-covariant position operator in 
relativistic quantum mechanics \cite{cab_Bacry1988}. We discuss these issues in
more detail in Sec.~\ref{sec:spin}.

In Ref.~\cite{CRW_2008_vector_bosons}, the EPR correlations in the spins of 
a pair of identical
relativistic spin-1 bosons were considered and the expectation values of various 
correlators calculated analytically. More recently a proposal was 
made~\cite{Barr2021} for testing Bell inequalities using a pair 
of spin-1 $W^+$ and $W^-$ bosons resulting from the decay of a spin-0 Higgs 
boson\footnote{Further analysis of the prospects for Bell violation measurements 
in bipartite systems of weak vector bosons at high-energy colliders have been 
described subsequently in
refs~\cite{ASBCM2022_entanglement_HtoZZ,APBW2022_entanglement-weak-decays,Aguilar-Saavedra2022_entanglement_HtoWW,FFGM2023}.}.
Several distinctive features of these recently-proposed measurements 
make them particularly interesting both in terms of their experimental setup and the
interpretation of the results. 
Firstly, the extremely short (sub-fm) sub-nuclear length scales of the bipartite systems 
under test are many orders of magnitude shorter than any existing measurements, 
making this a novel and unexplored regime. Secondly at least one $W$ must be
off-mass-shell if produced in a Higgs boson decay, since the mass of the Higgs boson 
is less than the sum of the $W^+$ and $W^-$ bosons. This opens the possibility of
performing Bell-inequality tests using rather virtual particles, again a regime in which we 
are not aware of existing tests. Finally, the use of “self-measuring” quantum spin 
states -- which exploit the chiral nature of the weak force to measure the bosons’ spins 
from the emitted direction of their daughter leptons -- challenges the assumption of 
the experimentalist’s freedom-of-choice that is typically made in Bell tests. 
These features provide motivation for increasing our understanding Bell-inequality 
violation in systems of vector bosons, however we do not rely on them in what follows.

\section{Scalar states of two vector bosons}

We are interested in states containing two relativistic vector bosons, which we label 
$a$ corresponding to the particle, and $b$ corresponding to its antiparticle. 
We consider here states that have sharp momentum, and label the four momentum 
of the particle $k$ and that of the antiparticle $p$. 
The simplifying assumption of states with sharp momentum allows the study of 
the relativistic effects without the additional complications associated with 
finite-width effects.  

The method of constructing the corresponding single-particle (and single-anti-particle)
states is presented for completeness in Appendices \ref{sec:app-field} 
and \ref{sec:app-states},
which themselves further develop the formalism described in
Ref.~\cite{CRW_2008_vector_bosons}.
The result is that a covariant boson-antiboson state corresponding to this situation takes 
the following form
\begin{equation}
e_{\lambda}^{\mu}(k) e_{\sigma}^{\nu}(p)
\ket{(k,\lambda);(p,\sigma)}
\equiv
e_{\lambda}^{\mu}(k) e_{\sigma}^{\nu}(p)
a^{\dagger}_{\lambda}(k) b^{\dagger}_{\sigma}(p) \ket{0},
\end{equation}
where the meanings of the amplitudes $e(k)$ and the creation operators 
$a^\dag, b^\dag$ are as described in Appendix \ref{sec:app-field}.

Here we consider a general scalar state in the following form
\begin{equation}
\ket{\alpha(k,p)} = 
g_{\mu\nu}(k,p) 
e_{\lambda}^{\mu}(k) e_{\sigma}^{\nu}(p)
\ket{(k,\lambda);(p,\sigma)},
\label{scalar-state-general}
\end{equation}
where
\begin{equation}
\label{g_def}
g_{\mu\nu}(k,p) =  \eta_{\mu\nu} 
+ \tfrac{c}{(kp)} \big( k_\mu p_\nu + p_\mu k_\nu \big),
\quad c \in{\mathbb{R}}.
\end{equation}
We note that transversality condition (\ref{e-transversality}) for amplitudes 
$e(k)$ reduces the second 
term in the bracket in (\ref{g_def}) to the $p_\mu k_\nu$ only.
Choosing the above parametrization we exclude from our
considerations the state 
$p_\mu k_\nu
e_{\lambda}^{\mu}(k) e_{\sigma}^{\nu}(p)
\ket{(k,\lambda);(p,\sigma)}$.
However, this state is separable.

The scalar state defined in Eq.~(\ref{scalar-state-general}) is normalized as follows
\begin{equation}
\bracket{\alpha(k,p)}{\alpha(k,p)} = 4 k^0 p^0 (\delta^3(\vec{0}))^2 A(k,p),
\end{equation}
with
\begin{equation}
A(k,p) = 2  +
\Big[ c\tfrac{m^2}{(kp)} - \tfrac{(kp)}{m^2} (1+ c) \Big]^2.
\end{equation}
Eq.~(\ref{scalar-state-general}) defines a whole variety of scalar states, 
two of which are distinguished. 

The first one, $\ket{\psi(k,p)}$,
corresponds to the choice $c_\psi=0$:
\begin{equation}
\ket{\psi(k,p)} = \eta_{\mu\nu} 
e_{\lambda}^{\mu}(k) e_{\sigma}^{\nu}(p)
\ket{(k,\lambda);(p,\sigma)}.
\label{scalar-state-psi}
\end{equation}
It is the simplest and the most natural scalar state. 
In \cite{CRW_2008_vector_bosons} we considered
the Einstein--Podolsky-Rosen type experiment with two bosons
in the state $\ket{\psi(k,p)}$.
The normalization factor $A(k,p)$ of the state $\ket{\psi(k,p)}$
has the simple form
\begin{equation}
A^\psi(k,p) = 2 +  \tfrac{(kp)^2}{m^4}.
\end{equation}

The second interesting scalar state,
$\ket{\xi(k,p)}$, corresponds to the choice 
$c_\xi = - 1$:
\begin{equation}
\ket{\xi(k,p)} = \Big( \eta_{\mu\nu} 
- \tfrac{1}{(kp)} p_\mu k_\nu \Big)
e_{\lambda}^{\mu}(k) e_{\sigma}^{\nu}(p)
\ket{(k,\lambda);(p,\sigma)}.
\label{scalar-state-xi}
\end{equation}
The state $\ket{\xi(k,p)}$ has the property that in the massless limit 
it converges to a scalar
two-photon state. This question was discussed in detail in 
\cite{Caban_2008_bosons_helicity}.
Notice also that the normalization factor $A(k,p)$ of the state
$\ket{\xi(k,p)}$ also takes the simple form
\begin{equation}
A^\xi(k,p) = 2 +  \tfrac{m^4}{(kp)^2}.
\end{equation}

\section{Spin operator for a relativistic particle}
\label{sec:spin}

When we want to calculate explicitly correlation functions, we need
to introduce the spin operator for relativistic massive particles.
The choice of such an operator is not a trivial problem.
We know that in the carrier space of a unitary representation of the Poincar\'e
group there exists a well-defined square of the spin operator
\begin{equation}
	\Hat{\Vec{S}}^2 =  s(s+1) I = - \frac{W^\mu W_\mu}{m^2} ,
	\label{eq:spin-square}
\end{equation}
where $s$ denotes spin of a particle and
\begin{equation}
	\label{Pauli_Lubanski}
	\hat{W}^{\mu}=
	\tfrac{1}{2}\epsilon^{\mu\nu\gamma\delta}\hat{P}_{\nu}\hat{J}_{\gamma\delta}
\end{equation}
is the Pauli-Lubanski four-vector, $\hat{P}_{\nu}$ is the
four-momentum operator, $\hat{J}_{\mu\nu}$ denote the generators of
the Lorentz group such that
$U(\Lambda)=\exp(i\omega^{\mu\nu}\hat{J}_{\mu\nu})$, and we assume
$\epsilon^{0123}=1$.
On the other hand, spin can be defined as a difference between total angular momentum 
and the orbital angular momentum 
$\Hat{\Vec{L}} = \Hat{\Vec{Q}}\times\Hat{\Vec{P}}$:
\begin{equation}
	\Hat{\Vec{S}} = \Hat{\Vec{J}} - \Hat{\Vec{Q}}\times\Hat{\Vec{P}}.
\end{equation}
Total angular momentum $\Hat{\Vec{J}}$ is well defined as the generator of the
rotations, $\Hat{J}^i = \varepsilon^{ijk} \Hat{J}^{jk}$, and the momentum operator
$\Hat{\Vec{P}}$ is also well defined (compare Eq.~(\ref{four-momentum})),
but there does not exist a generally accepted position operator
$\Hat{\Vec{Q}}$\cite{cab_Bacry1988}.
Different choices of the position operator $\Hat{\Vec{Q}}$ lead to different spin
operators. The most popular position operator was introduced by 
Newton and Wigner\cite{cab_NW1949}
\begin{equation}
	\Hat{\Vec{Q}}_{NW} =  - \frac{1}{2} \Big[
	\frac{1}{\hat{P}^0}\hat{\vec{K}} + \hat{\vec{K}}
	\frac{1}{\hat{P}^0} \Big] - 
	\frac{\hat{\vec{P}} \times \hat{\vec{W}}}{m \hat{P}^0 (m + \hat{P}^0 )}, 
	\label{NW-operator}
\end{equation}
where $\Hat{\Vec{K}}$ denotes the boost generators, $\vec{K}^i = \vec{J}^{0i}$.
The Newton-Wigner position operator possesses many desirable properties:
it is a vector with commuting, self-adjoint components and it is defined for
arbitrary spin. Unfortunately, $\Hat{\Vec{Q}}_{NW}$ does not transform 
in a manifestly covariant way under Lorentz boosts. 

The spin operator related to the Newton-Wigner
position operator is equal to
\begin{equation}
	\label{spin}
	\hat{\vec{S}}=
	\frac{1}{m}
	\Big(\hat{\vec{W}}-\hat{W}^0\frac{\Hat{\Vec{P}}}{\hat{P}^0+m}\Big).
\end{equation}
This operator has several desirable features.
Components of this operator satisfy the standard su(2) Lie algebra commutation relations. Moreover, the square of the operator (\ref{spin}) in 
a unitary irreducible representation of the Poincar\'e group is equal to
(\ref{eq:spin-square}). What more, the operator (\ref{spin}) is the only axial vector
which is a linear function of the Pauli-Lubanski four-vector components\cite{cab_BLT1969}.
Finally, as was shown in \cite{CRW_2013_Dirac_spin}, the operator $\hat{\vec{S}}$
has an elegant transformation formula 
$\hat{\vec{S}}^\prime = R (\Lambda,\hat{P}) \hat{\vec{S}}$
under Lorentz group action,
where $R (\Lambda,\hat{P})$ is the corresponding Wigner rotation 
while $\hat{P}$ is the four-momentum operator.
That is why in our opinion the spin operator (\ref{spin}) is the most appropriate 
one and we chose it for our calculations (compare \cite{CRW_2009_strange_correl}).

The influence of the Newton-Wigner localization of a particle inside a detector 
during spin measurement on relativistic quantum correlations,
but in a case of a fermion pair,
was considered in \cite{CRRSW2014_correl-localization}.
In Appendix \ref{sec:localization} we have shortly recalled
these results adapting them to a system of vector bosons.
We concluded there that for real, massive particles there is no problem
with localization of a particle inside the detector in situation in which it is 
directly detected. The extent to which such arguments
might also be relevant to measurements of gauge boson
spin via the kinematics of their chiral decays, as was used
e.g. in \cite{Barr2021}, is a more involved question, and lies beyond
the scope of the present paper.

We note that other spin operators have been also used in the description
of relativistic EPR experiments, the most popular one is the operator used
by Czachor\cite{cab_Czachor1997_1}. This operator is related with the so-called center
of mass position operator which has non-commuting components.
For more exhaustive discussion of the problem of choice of the proper relativistic
spin operator see,
e.g.,~\cite{Terno2003,cab_CR2005, CR2006,CRW_2009_strange_correl,SV_2012_Wigner_rotation_spin,BAKG2014_Spin,CKT2016_PhysRevA.94.062115,ZZS2020_position-spin-rel-QM,TA2021_relativistic-spin,Lee2022}.

The spin operator $\hat{\vec{S}}$ (\ref{spin}) acts on one-particle
states according to
\begin{equation}
	\label{Spin_action_basis}
	\hat{\vec{S}}\ket{k,\sigma}=\vec{S}_{\lambda\sigma}\ket{k,\lambda},
\end{equation}
where $S^i$ are standard spin-1 matrices (compare, e.g., \cite{cab_Ballentine2014}):
\begin{gather}
	\label{spin_1}
	S^1=\tfrac{1}{\sqrt{2}}
	\begin{pmatrix}
		0 & 1 & 0 \\
		1 & 0 & 1 \\
		0 & 1 & 0 \\
	\end{pmatrix},\quad
	S^2=\tfrac{i}{\sqrt{2}}
	\begin{pmatrix}
		0 & -1 & 0 \\
		1 & 0 & -1 \\
		0 & 1 & 0 \\
	\end{pmatrix},\\
	S^3=
	\begin{pmatrix}
		1 & 0 & 0 \\
		0 & 0 & 0 \\
		0 & 0 & -1 \\
	\end{pmatrix}.
\end{gather}
An operator which acts like a spin on particles whose momenta belong to some 
definite region $\Omega$ in momentum space and gives 0 otherwise
has the form
\begin{equation}
	\Hat{\vec{S}}^{a}_{\Omega} = 
	\int_{\Omega} \tfrac{d^3 \vec{k}}{2 k^0} a^{\dagger}(k) \vec{S} a(k),
\end{equation}
where $a(k)=\big(a_{+1}(k),a_{0}(k),a_{-1}(k)\big)^T$.
A similar operator but acting on antiparticles has the form
\begin{equation}
	\Hat{\vec{S}}^{b}_{\Omega} = 
	\int_{\Omega} \tfrac{d^3 \vec{k}}{2 k^0} b^{\dagger}(k) \vec{S} b(k).
\end{equation}
We have
\begin{equation}
	\Hat{\vec{S}}^{a}_{\Omega} \ket{(k,\lambda)_a;(p,\sigma)_b} = 
	\chi_\Omega(k) \vec{S}_{\lambda^\prime \lambda}
	\ket{(k,\lambda^\prime)_a;(p,\sigma)_b} 
\end{equation}
and
\begin{equation}
	\Hat{\vec{S}}^{b}_{\Omega} \ket{(k,\lambda)_a;(p,\sigma)_b} = 
	\chi_\Omega(p) \vec{S}_{\sigma^\prime \sigma}
	\ket{(k,\lambda)_a;(p,\sigma^\prime)_b},
\end{equation}
where $\chi_\Omega$ denotes the characteristic function of the set
$\Omega$: $\chi_\Omega(k)=1$ for $k\in\Omega$ and 
$\chi_\Omega(k)=0$ for $k\not\in\Omega$.
The spectral decomposition of the operator 
$\vecg{\omega} \cdot \Hat{\vec{S}}^{a}_{\Omega}$
can be written as
\begin{equation}
	\vecg{\omega} \cdot \Hat{\vec{S}}^{a}_{\Omega} = 
	1\cdot \Pi^{a+}_{\Omega\vecg{\omega}}
	+ (-1)\cdot \Pi^{a-}_{\Omega\vecg{\omega}}
	+ 0\cdot \Pi^{a0}_{\Omega\vecg{\omega}},
\end{equation}
where the projectors act in the following way:
\begin{multline}
	\Pi^{a\pm}_{\Omega\vecg{\omega}} 
	\ket{(k,\lambda)_a;(p,\sigma)_b} = \\
	\tfrac{1}{2} \chi_\Omega(k)
	\Big(
	(\vecg{\omega}\cdot\vec{S})^{2}_{\lambda^\prime \lambda}
	\pm
	(\vecg{\omega}\cdot\vec{S})_{\lambda^\prime \lambda}
	\Big) 
	\ket{(k,\lambda^\prime)_a;(p,\sigma)_b},
	\label{projector-pm}
\end{multline}
\begin{multline}
	\Pi^{a0}_{\Omega\vecg{\omega}} 
	\ket{(k,\lambda)_a;(p,\sigma)_b} = \\
	\chi_\Omega(k)
	\Big(
	\delta_{\lambda^\prime \lambda} -
	(\vecg{\omega}\cdot\vec{S})^{2}_{\lambda^\prime \lambda}
	\Big) 
	\ket{(k,\lambda^\prime)_a;(p,\sigma)_b}.
	\label{projector-0}
\end{multline}
Analogous formulas can easily be found for 
$\vecg{\omega} \cdot \Hat{\vec{S}}^{b}_{\Omega}$.

\section{Probabilities}

Now, let two distant observers, Alice and Bob, be at rest with respect to 
a given inertial frame and share a pair of bosons in the scalar state 
$\ket{\alpha(k,p)}$ (Eq.~(\ref{scalar-state-general})).
The probability that Alice obtains $\sigma$ and Bob $\lambda$ 
($\lambda,\sigma\in\{-1,0,1\}$),
when measuring spin projections
on the directions $\vec{a}$ and $\vec{b}$, respectively,
are given by the formula
\begin{equation}
\label{probabilities_general_def}
P_{\sigma\lambda} = 
\frac{\bra{\alpha(k,p)}
\hat{\Pi}_{A\vec{a}}^{\sigma} \hat{\Pi}_{B\vec{b}}^{\lambda}
\ket{\alpha(k,p)}}{\bracket{\alpha(k,p)}{\alpha(k,p)}}.
\end{equation}
Here we assume that Alice (Bob) can register only particles 
(antiparticles) whose momenta 
belong to the region $A$ ($B$) in the momentum space
and that they use the spin operator given in Eq.~(\ref{spin}).
Projectors in the above formula are defined in 
Eqs.~(\ref{projector-pm},\ref{projector-0}).
Further, assuming that Alice can measure only bosons with
four-momentum $k$ and Bob those with four-momentum $p$ we find
\begin{subequations}
\label{probabilities-general}
\begin{multline}
P_{\pm\pm}  =  \frac{1}{4A(k,p)}
\tr \{M(k,\vec{a}) g(k,p) M(p,\vec{b}) g(k,p)^T \\
- N(k,\vec{a}) g(k,p) N(p,\vec{b}) g(k,p)^T\},
\end{multline}
\begin{multline}
P_{\pm\mp}  = \frac{1}{4A(k,p)}
\tr \{M(k,\vec{a}) g(k,p) M(p,\vec{b}) g(k,p)^T  \\
+ N(k,\vec{a}) g(k,p) N(p,\vec{b}) g(k,p)^T\},
\end{multline}
\begin{equation}
P_{0\pm}  =  \frac{1}{2A(k,p)}
\tr \{T(k,\vec{a}) g(k,p) M(p,\vec{b}) g(k,p)^T\},
\end{equation}
\begin{equation}
P_{\pm 0}  =  \frac{1}{2A(k,p)}
\tr \{M(k,\vec{a}) g(k,p) T(p,\vec{b}) g(k,p)^T\},
\end{equation}
\begin{equation}
P_{00}  =  \frac{1}{A(k,p)}
\tr \{T(k,\vec{a}) g(k,p) T(p,\vec{b}) g(k,p)^T\},
\end{equation}
\end{subequations}
where
\begin{equation}
g(k,p) = [g_{\mu\nu}(k,p)]
\end{equation}
and where we have introduced the following notation
\begin{equation}
M(q,\vecg{\omega}) = [M(q,\vecg{\omega})^{\mu\nu}]
 = e^{*\mu}_{\lambda}(q) 
(\vecg{\omega}\cdot\vec{S})^{2}_{\lambda\sigma}
 e^{\nu}_{\sigma}(q),
\end{equation}
\begin{equation}
N(q,\vecg{\omega}) = [N(q,\vecg{\omega})^{\mu\nu}]
= e^{*\mu}_{\lambda}(q) 
(\vecg{\omega}\cdot\vec{S})_{\lambda\sigma}
e^{\nu}_{\sigma}(q),
\end{equation}
\begin{equation}
T(q,\vecg{\omega}) = [T(q,\vecg{\omega})^{\mu\nu}]
= e^{*\mu}_{\lambda}(q) 
\big( \delta_{\lambda\sigma}-(\vecg{\omega}\cdot\vec{S})^{2}_{\lambda\sigma}\big)
e^{\nu}_{\sigma}(q).
\end{equation}

With the help of the above formulas one can find $P_{\lambda\sigma}$
for arbitrary $k$ and $p$. However, the resulting formulas are complicated.
Thus, here we restrict ourselves to the situation when Alice and Bob's frame
coincides with the center of mass frame of the boson pair ($p=k^\pi=(k^0,-\vec{k})$).
Let us introduce the notation
\begin{equation}
x=\tfrac{\vec{k}^2}{m^2},\quad 
\vec{n} = \tfrac{\vec{k}}{|\vec{k}|}.
\label{notation_x}
\end{equation}
Using this notation we have in the center of mass frame
\begin{equation}
A(k,k^\pi) = 2+ \tfrac{[4x(x+1)(c+1)+1]^2}{(2x+1)^2}
\end{equation}
and
\begin{subequations}
\label{probabilities-center-of-mass}
\begin{widetext}
\begin{multline}
P_{\pm\pm} = \frac{1}{4A(k,k^\pi)}
\Big\{ 
1 + 4x(x+1) \big(1+2c\tfrac{x}{2x+1}\big) \big( 1+2c\tfrac{x+1}{2x+1}\big)
\big[ 1- (\vec{a}\cdot\vec{n})^2 - (\vec{b}\cdot\vec{n})^2 \big]
+ \Big[
(\vec{a}\cdot\vec{b}) 
+ 2x\big( 1+2c\tfrac{x+1}{2x+1}\big) (\vec{a}\cdot\vec{n}) (\vec{b}\cdot\vec{n})
\Big]^2
\\
- 2
\Big[
(\vec{a}\cdot\vec{b}) + 2x \big( 1+2c\tfrac{x+1}{2x+1}\big)
\big[ (\vec{a}\cdot\vec{b}) - (\vec{a}\cdot\vec{n}) (\vec{b}\cdot\vec{n})
\big]
\Big]
\Big\},
\end{multline}	
\begin{multline}
P_{\pm\mp} = \frac{1}{4A(k,k^\pi)}
\Big\{ 
1 + 4x(x+1) \big(1+2c\tfrac{x}{2x+1}\big) \big( 1+2c\tfrac{x+1}{2x+1}\big)
\big[ 1- (\vec{a}\cdot\vec{n})^2 - (\vec{b}\cdot\vec{n})^2 \big]
+ \Big[
(\vec{a}\cdot\vec{b}) 
+ 2x\big( 1+2c\tfrac{x+1}{2x+1}\big) (\vec{a}\cdot\vec{n}) (\vec{b}\cdot\vec{n})
\Big]^2
\\
+ 2
\Big[
(\vec{a}\cdot\vec{b}) + 2x \big( 1+2c\tfrac{x+1}{2x+1}\big)
\big[ (\vec{a}\cdot\vec{b}) - (\vec{a}\cdot\vec{n}) (\vec{b}\cdot\vec{n})
\big]
\Big]
\Big\},
\end{multline}
\begin{multline}
P_{0\pm} = \frac{1}{2A(k,k^\pi)}
\Big\{ 
1 + 4x(x+1)\big(1+2c\tfrac{x}{2x+1}\big) \big( 1+2c\tfrac{x+1}{2x+1}\big)
 (\vec{a}\cdot\vec{n})^2 
- \Big[
(\vec{a}\cdot\vec{b}) 
+ 2x\big( 1+2c\tfrac{x+1}{2x+1}\big) (\vec{a}\cdot\vec{n}) (\vec{b}\cdot\vec{n})
\Big]^2 
\Big\},
\end{multline}
\begin{multline}
P_{\pm 0} = \frac{1}{2A(k,k^\pi)}
\Big\{ 
1 + 4x(x+1) \big(1+2c\tfrac{x}{2x+1}\big) \big( 1+2c\tfrac{x+1}{2x+1}\big) (\vec{b}\cdot\vec{n})^2
- \Big[
(\vec{a}\cdot\vec{b}) 
+ 2x\big( 1+2c\tfrac{x+1}{2x+1}\big) (\vec{a}\cdot\vec{n}) (\vec{b}\cdot\vec{n})
\Big]^2 
\Big\},
\end{multline}
\begin{equation}
P_{00} = \frac{1}{A(k,k^\pi)}
\Big[
(\vec{a}\cdot\vec{b})
+ 2x\big( 1+2c\tfrac{x+1}{2x+1}\big) (\vec{a}\cdot\vec{n}) (\vec{b}\cdot\vec{n})
\Big]^2.
\end{equation}
\newpage
\end{widetext}
\end{subequations}

The correlation function defined as
\begin{align}
C(\vec{a},k;\vec{b},k^\pi) & = \sum_{\lambda,\sigma=-1,0,1} \lambda \sigma P_{\lambda\sigma} 
 = 2(P_{\pm\pm}-P_{\pm\mp})
\end{align}
is equal to
\begin{multline}
C(\vec{a},k;\vec{b},k^\pi) = - \tfrac{2}{A(k,k^\pi)}
\Big[
(\vec{a}\cdot\vec{b})\\
 + 2x \big( 1 + 2c\tfrac{x+1}{2x+1}\big)
\Big( (\vec{a}\cdot\vec{b}) - (\vec{a}\cdot\vec{n}) (\vec{b}\cdot\vec{n})
\Big)
\Big].
\label{correl_function_general}
\end{multline}
Let us now consider the nonrelativistic and ultrarelativistic limits of the above probabilities.

\subsubsection{Nonrelativistic limit}

In the nonrelativistic limit ($x\to 0$) we obtain
\begin{subequations}
\begin{equation}
P_{\pm\pm}(x\to0) = \tfrac{1}{12}[1-(\vec{a}\cdot\vec{b})]^2,
\end{equation}
\begin{equation}
P_{\pm\mp}(x\to0) = \tfrac{1}{12}[1+(\vec{a}\cdot\vec{b})]^2,
\end{equation}
\begin{equation}
P_{0\pm}(x\to0) = P_{\pm0}(x\to0) = \tfrac{1}{6}[1-(\vec{a}\cdot\vec{b})^2],
\end{equation}
\begin{equation}
P_{00}(x\to0) = \tfrac{1}{3}(\vec{a}\cdot\vec{b})^2.
\end{equation}
\end{subequations}
These probabilities
are the same as those calculated in the framework of nonrelativistic quantum
mechanics in the singlet spin-1 state
\begin{equation}
\ket{\Psi^{\mathrm{nonrel}}_{\mathrm{singlet}}} = 
\tfrac{1}{\sqrt{3}} 
\big(
\ket{1}\ket{-1} - \ket{0}\ket{0} + \ket{-1}\ket{1}
\big).
\label{nonrel_singlet}
\end{equation}

\subsubsection{Ultrarelativistic limit}

In the ultrarelativistic limit ($x\to\infty$) we have to distinguish two separate cases:
$c=-1$, corresponding to the state $\ket{\xi}$, and $c\not=-1$ which contains 
the state $\ket{\psi}$.

In the case $c\not=-1$ we get
\begin{subequations}
\begin{multline}
P^{c\not=-1}_{\pm\pm}(x\to\infty) = P^{c\not=-1}_{\pm\mp}(x\to\infty)\\
= 
\tfrac{1}{4} 
\big[ 1-(\vec{a}\cdot\vec{n})^2 \big]
\big[ 1-(\vec{b}\cdot\vec{n})^2 \big],
\end{multline}
\begin{equation}
P^{c\not=-1}_{0\pm}(x\to\infty)=
\tfrac{1}{2} 
(\vec{a}\cdot\vec{n})^2 
\big[ 1-(\vec{b}\cdot\vec{n})^2 \big],
\end{equation}
\begin{equation}
P^{c\not=-1}_{\pm 0}(x\to\infty)=
\tfrac{1}{2} 
(\vec{b}\cdot\vec{n})^2 
\big[ 1-(\vec{a}\cdot\vec{n})^2 \big],
\end{equation}
\begin{equation}
P^{c\not=-1}_{00}(x\to\infty)=
(\vec{a}\cdot\vec{n})^2 
(\vec{b}\cdot\vec{n})^2.
\end{equation}
\end{subequations}
The correlation function in this case vanishes
\begin{equation}
C^{c\not=-1}(\vec{a},k;\vec{b},k^\pi,x\to\infty) = 0.
\end{equation}
The case $c\not=-1$ includes the state $\ket{\psi(k,k^\pi)}$.
The probabilities in the ultrarelativistic limit in this state 
were given in Eq.~(60) in Ref.~\cite{CRW_2008_vector_bosons}
and they coincide with the above formulas.

The case $c=-1$ corresponds to the state $\ket{\xi(k,k^\pi)}$ and for this state
we obtain
\begin{subequations}
\begin{multline}
P^{\xi}_{\pm\pm}(x\to\infty) = \tfrac{1}{8} 
\big\{
\big[
(\vec{a}\cdot\vec{b}) - (\vec{a}\cdot\vec{n}) (\vec{b}\cdot\vec{n})
\big]^2\\
+
\big[
(\vec{a}\cdot\vec{n}) - (\vec{b}\cdot\vec{n})
\big]^2
\big\},
\end{multline}
\begin{multline}
P^{\xi}_{\pm\mp}(x\to\infty) = \tfrac{1}{8} 
\big\{
\big[
(\vec{a}\cdot\vec{b}) - (\vec{a}\cdot\vec{n}) (\vec{b}\cdot\vec{n})
\big]^2\\
+
\big[
(\vec{a}\cdot\vec{n}) + (\vec{b}\cdot\vec{n})
\big]^2
\big\},
\end{multline}
\begin{multline}
P^{\xi}_{0\mp}(x\to\infty) = \tfrac{1}{4} 
\big\{
1 - (\vec{a}\cdot\vec{n})^2\\
- \big[
(\vec{a}\cdot\vec{b}) - (\vec{a}\cdot\vec{n}) (\vec{b}\cdot\vec{n})
\big]^2
\big\},
\end{multline}
\begin{multline}
P^{\xi}_{\mp0}(x\to\infty) = \tfrac{1}{4} 
\big\{
1 - (\vec{b}\cdot\vec{n})^2\\
- \big[
(\vec{a}\cdot\vec{b}) - (\vec{a}\cdot\vec{n}) (\vec{b}\cdot\vec{n})
\big]^2
\big\},
\end{multline}
\begin{equation}
P^{\xi}_{00}(x\to\infty) = \tfrac{1}{2} 
\big[
(\vec{a}\cdot\vec{b}) - (\vec{a}\cdot\vec{n}) (\vec{b}\cdot\vec{n})
\big]^2,
\end{equation}
\end{subequations}
while the correlation function is equal to
\begin{equation}
C^\xi(\vec{a},k;\vec{b},k^\pi,x\to\infty) = 
- (\vec{a}\cdot\vec{n})(\vec{b}\cdot\vec{n}).
\end{equation}
It is interesting to observe that the state $\ket{\xi(k,k^\pi)}$ is distinguished  
by the property that for this state only the correlation function does not vanish
in the ultrarelativistic limit.

In the following we concentrate on the most interesting (and natural) states $\ket{\psi}$ 
and $\ket{\xi}$.

\subsubsection{Probabilities in the state $|\psi(k,k^\pi)\rangle$}

We have calculated these probabilities in our previous paper \cite{CRW_2008_vector_bosons},
they are given by Eq.~(58) from \cite{CRW_2008_vector_bosons}.
For convenience we recall them here
\begin{subequations}
\begin{multline}
P^{\psi}_{\pm\pm} = \tfrac{1}{4[(2x+1)^2+2]}
\big\{
(2x+1)^2 -2(2x+1) (\vec{a}\cdot\vec{b})\\
+ 4x (\vec{a}\cdot\vec{n}) (\vec{b}\cdot\vec{n})
- 4x(x+1) \big[
(\vec{a}\cdot\vec{n})^2 + (\vec{b}\cdot\vec{n})^2
\big]\\
+ \big[
(\vec{a}\cdot\vec{b}) + 2x (\vec{a}\cdot\vec{n}) (\vec{b}\cdot\vec{n})
\big]^2
\big\},
\end{multline}
\begin{multline}
P^{\psi}_{\pm\mp} = \tfrac{1}{4[(2x+1)^2+2]}
\big\{
(2x+1)^2 + 2(2x+1) (\vec{a}\cdot\vec{b})\\
- 4x (\vec{a}\cdot\vec{n}) (\vec{b}\cdot\vec{n})
- 4x(x+1) \big[
(\vec{a}\cdot\vec{n})^2 + (\vec{b}\cdot\vec{n})^2
\big]\\
+ \big[
(\vec{a}\cdot\vec{b}) + 2x (\vec{a}\cdot\vec{n}) (\vec{b}\cdot\vec{n})
\big]^2
\big\},
\end{multline}
\begin{multline}
P^{\psi}_{0\mp} = \tfrac{1}{2[(2x+1)^2+2]}
\big\{
1 + 4x(x+1) (\vec{a}\cdot\vec{n})^2\\
- \big[
(\vec{a}\cdot\vec{b}) + 2x (\vec{a}\cdot\vec{n}) (\vec{b}\cdot\vec{n})
\big]^2
\big\},
\end{multline}
\begin{multline}
P^{\psi}_{\mp0} = \tfrac{1}{2[(2x+1)^2+2]}
\big\{
1 + 4x(x+1) (\vec{b}\cdot\vec{n})^2\\
- \big[
(\vec{a}\cdot\vec{b}) + 2x (\vec{a}\cdot\vec{n}) (\vec{b}\cdot\vec{n})
\big]^2
\big\},
\end{multline}
\begin{equation}
P^{\psi}_{00} = \tfrac{1}{(2x+1)^2+2}
\big[
(\vec{a}\cdot\vec{b}) + 2x (\vec{a}\cdot\vec{n}) (\vec{b}\cdot\vec{n})
\big]^2.
\end{equation}
\end{subequations}
The correlation function for the state $\ket{\psi(k,k^\pi)}$ was given in Eq.~(59) 
in \cite{CRW_2008_vector_bosons} and reads
\begin{multline}
C^{\psi}(\vec{a},k;\vec{b},k^\pi) = \tfrac{- 2}{(2x+1)^2+2}
\big[
(2x+1) (\vec{a}\cdot\vec{b})\\
 - 2x (\vec{a}\cdot\vec{n})(\vec{b}\cdot\vec{n})
\big].
\end{multline}

\subsubsection{Probabilities in the state $|\xi(k,k^\pi)\rangle$}

In this case we insert into Eq.~(\ref{probabilities-center-of-mass}) the following
values:
\begin{equation}
c_\xi = - 1, \quad
A^\xi(k,k^\pi)=2 + \tfrac{1}{(2x+1)^2}.
\end{equation}
The resulting probabilities have the following form
\begin{subequations}
\begin{multline}
P^{\xi}_{\pm\pm} = 
\tfrac{1}{4[2(2x+1)^2+1]}
\big\{
\big[
(2x+1) (\vec{a}\cdot\vec{b}) -2x (\vec{a}\cdot\vec{n})(\vec{b}\cdot\vec{n})
\big]^2\\
+ (2x+1)^2 - 4x(x+1)\big[1 - (\vec{a}\cdot\vec{n})^2 - (\vec{b}\cdot\vec{n})^2\big]\\
- 2(2x+1) \big[(\vec{a}\cdot\vec{b}) 
+ 2x (\vec{a}\cdot\vec{n})(\vec{b}\cdot\vec{n})\big]
\big\},
\end{multline}
\begin{multline}
P^{\xi}_{\pm\mp} = 
\tfrac{1}{4[2(2x+1)^2+1]}
\big\{
\big[
(2x+1) (\vec{a}\cdot\vec{b}) -2x (\vec{a}\cdot\vec{n})(\vec{b}\cdot\vec{n})
\big]^2\\
+ (2x+1)^2 - 4x(x+1)\big[1 - (\vec{a}\cdot\vec{n})^2 - (\vec{b}\cdot\vec{n})^2\big]\\
+ 2(2x+1) \big[(\vec{a}\cdot\vec{b}) 
+ 2x (\vec{a}\cdot\vec{n})(\vec{b}\cdot\vec{n})\big]
\big\},
\end{multline}
\begin{multline}
P^{\xi}_{0\pm} = 
\tfrac{1}{2[2(2x+1)^2+1]}
\big\{
1 +  4x(x+1)\big[1 - (\vec{a}\cdot\vec{n})^2\big]\\
- \big[
(2x+1) (\vec{a}\cdot\vec{b}) -2x (\vec{a}\cdot\vec{n})(\vec{b}\cdot\vec{n})
\big]^2
\big\},
\end{multline}
\begin{multline}
P^{\xi}_{\pm 0} = 
\tfrac{1}{2[2(2x+1)^2+1]}
\big\{
1 +  4x(x+1)\big[1 - (\vec{b}\cdot\vec{n})^2\big]\\
- \big[
(2x+1) (\vec{a}\cdot\vec{b}) -2x (\vec{a}\cdot\vec{n})(\vec{b}\cdot\vec{n})
\big]^2
\big\},
\end{multline}
\begin{multline}
P^{\xi}_{00} = 
\tfrac{1}{2(2x+1)^2+1}
\big[
(2x+1) (\vec{a}\cdot\vec{b}) -2x (\vec{a}\cdot\vec{n})(\vec{b}\cdot\vec{n})
\big]^2.
\end{multline}
\end{subequations}
The correlation function (\ref{correl_function_general})
in the state $\ket{\xi(k,k^\pi)}$ reads
\begin{equation}
C^\xi(\vec{a},k;\vec{b},k^\pi) = - \tfrac{2(2x+1)}{2(2x+1)^2+1} 
\big[
(\vec{a}\cdot\vec{b}) + 2x (\vec{a}\cdot\vec{n})(\vec{b}\cdot\vec{n})
\big].
\end{equation}

\section{Bell-type inequalities}

Now we are in a position to discuss the violation of Bell-type inequalities in a system of 
two vector bosons. We restrict our considerations to the situation when Alice and Bob
are in the center--of--mass frame of a boson pair. We consider here three inequalities:
the Clauser--Horne--Shimony--Holt (CHSH) inequality \cite{cab_CHSH1969}, 
the Mermin inequality \cite{cab_Mermin1980}
and the Collins--Gisin--Linden--Massar--Popescu (CGLMP) inequality
\cite{CGLMP_Bell_ineq_high_spin}.

\paragraph*{CHSH inequality.}
The generalized CHSH inequality can be written in the form
\begin{equation}
|C(\vec{a},\vec{b}) - C(\vec{a},\vec{d})| +
|C(\vec{c},\vec{b}) + C(\vec{c},\vec{d})| \le 2,
\end{equation}
where $C(\vec{a},\vec{b})$ is the correlation function of spin projections
on the directions $\vec{a}$ and $\vec{b}$.
The CHSH inequality is optimal for detecting quantum nonlocality
in a system of two qubits.
However, in a system of two spin 1 particles in a singlet state
in nonrelativistic quantum mechanics
the CHSH inequality cannot be violated.
In our paper \cite{CRW_2008_vector_bosons} we have considered the violation of 
the CHSH inequality in the state $\ket{\psi(k,k^\pi)}$,
we have shown that this inequality is not violated in $\ket{\psi(k,k^\pi)}$.
Our further numerical simulations also show that the CHSH inequality cannot be violated 
in the state $\ket{\xi(k,k^\pi)}$, either.

\begin{figure}
	\includegraphics[width=0.95\columnwidth]{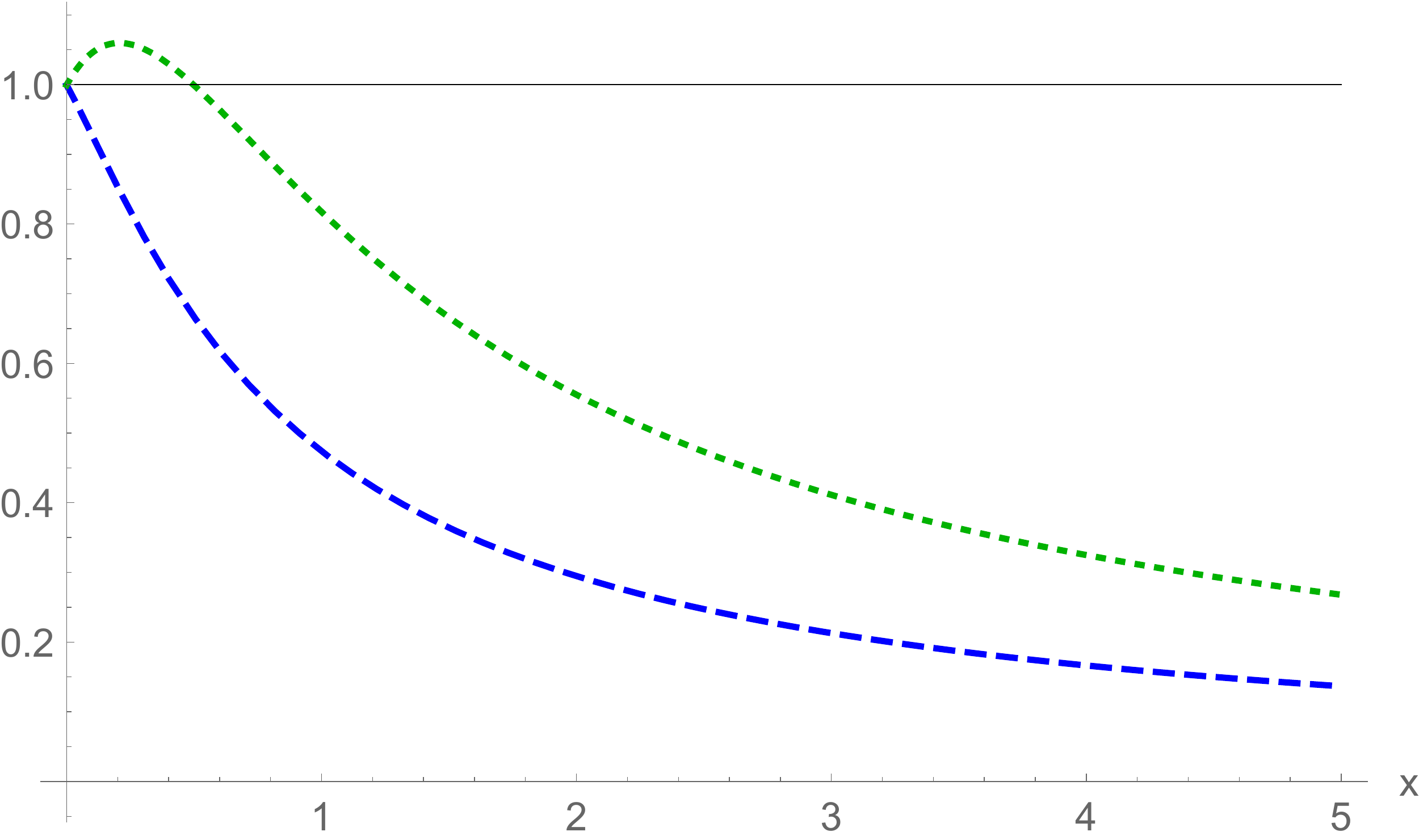}
	\caption{Comparison of the violation of the Mermin inequality in the 
		state $\ket{\xi(k,k^\pi)}$ (blue, dashed line) and in the state 
		$\ket{\psi(k,k^\pi)}$ (green, dotted line). The configuration of particles momenta
		and measurements directions is the following:
		$\vec{n}=(0,0,1)$, 
		$\vec{w}=(\cos\phi_w \sin\theta_w, \sin\phi_w \sin\theta_w, \cos\theta_w)$,
		$\vec{w}\in\{\vec{a},\vec{b},\vec{c}\}$ and
		$\theta_a= 1.593$, $\phi_a= 3.236$,
		$\theta_b= 1.564$, $\phi_b= 1.150$,
		$\theta_c= 1.514$, $\phi_c= 5.322$.}
	\label{fig:Mermin_xi_psi-1}
\end{figure}

\begin{figure}
	\includegraphics[width=0.95\columnwidth]{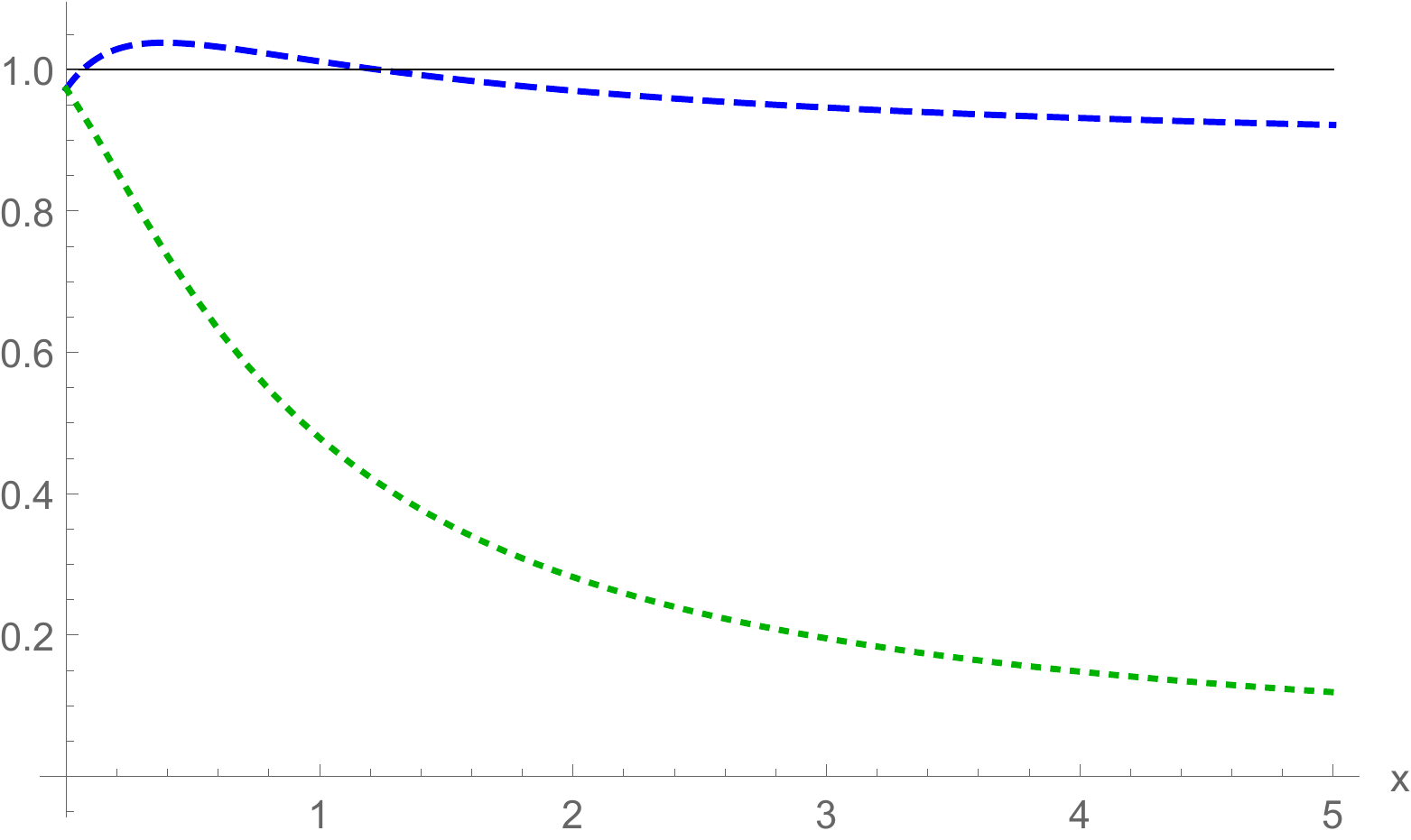}
	\caption{Comparison of the violation of the Mermin inequality in the 
		state $\ket{\xi(k,k^\pi)}$ (blue, dashed line) and in the state 
		$\ket{\psi(k,k^\pi)}$ (green, dotted line). The configuration of particles momenta
		and measurements directions is the following:
		$\vec{n}=(0,0,1)$, 
		$\vec{w}=(\cos\phi_w \sin\theta_w, \sin\phi_w \sin\theta_w, \cos\theta_w)$,
		$\vec{w}\in\{\vec{a},\vec{b},\vec{c}\}$ and
		$\theta_a= 1.891$, $\phi_a= 3.820$,
		$\theta_b= 2.589$, $\phi_b= 0.653$,
		$\theta_c= 0.220$, $\phi_c= 0.716$.}
	\label{fig:Mermin_xi_psi-2}
\end{figure}

\paragraph*{Mermin inequality.}
The Mermin inequality for spin 1 particles reads
\begin{equation}
C(\vec{a},\vec{b}) + C(\vec{b},\vec{c}) + C(\vec{c},\vec{a}) \le 1.
\end{equation}
As it was shown in \cite{cab_Mermin1980} this inequality should be satisfied
in any local, realistic theory. 
This inequality cannot be violated in nonrelativistic quantum mechanics.
However, as we have shown in \cite{CRW_2008_vector_bosons}, relativistic
vector bosons in the state $\ket{\psi(k,k^\pi)}$ can violate the Mermin inequality.
We find that bosons in the state $\ket{\xi(k,k^\pi)}$ also can violate 
the Mermin inequality.
In Figs.~\ref{fig:Mermin_xi_psi-1},\ref{fig:Mermin_xi_psi-2} we have compared 
the violation of the Mermin inequality in the states $\ket{\psi(k,k^\pi)}$ 
and $\ket{\xi(k,k^\pi)}$ in different configurations.

\begin{figure}
\includegraphics[width=0.95\columnwidth]{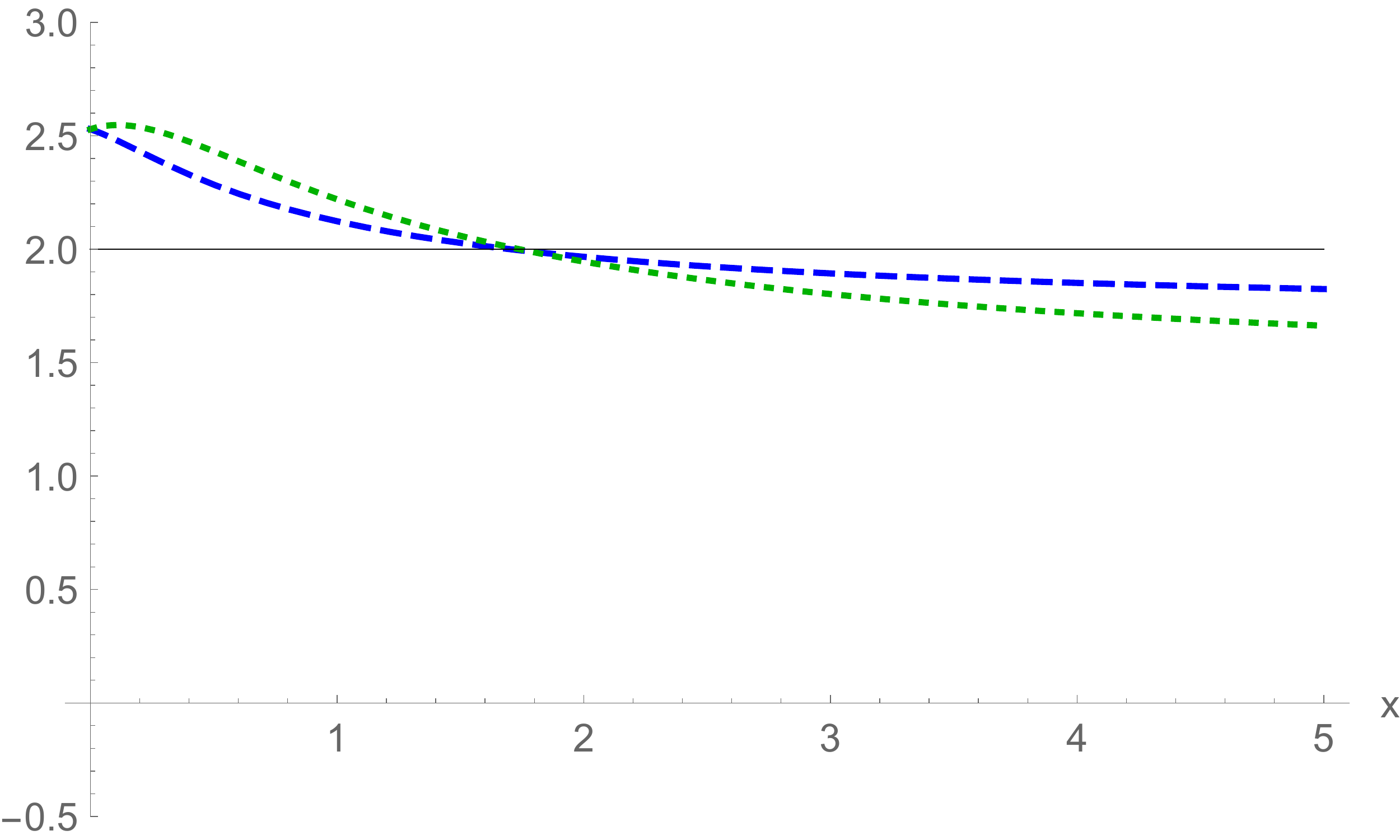}
\caption{Comparison of the violation of the CGLMP inequality in the 
	state $\ket{\xi(k,k^\pi)}$ (blue, dashed line) and in the state 
    $\ket{\psi(k,k^\pi)}$ (green, dotted line). The configuration of particles momenta
    and measurements directions is the following:
	$\vec{n}=(0,0,1)$, 
	$\vec{w}=(\cos\phi_w \sin\theta_w, \sin\phi_w \sin\theta_w, \cos\theta_w)$,
	$\vec{w}\in\{\vec{a},\vec{b},\vec{c},\vec{d}\}$ and
	$\theta_a= 2.667$, $\phi_a= 4.109$,
	$\theta_b= 0.924$, $\phi_b= 0.974$,
	$\theta_c= 2.699$, $\phi_c= 1.005$,
	$\theta_d= 0$, $\phi_d= 0$.} 
\label{fig:CGLMP_xi_psi-1}
\end{figure}

\begin{figure}
\includegraphics[width=0.95\columnwidth]{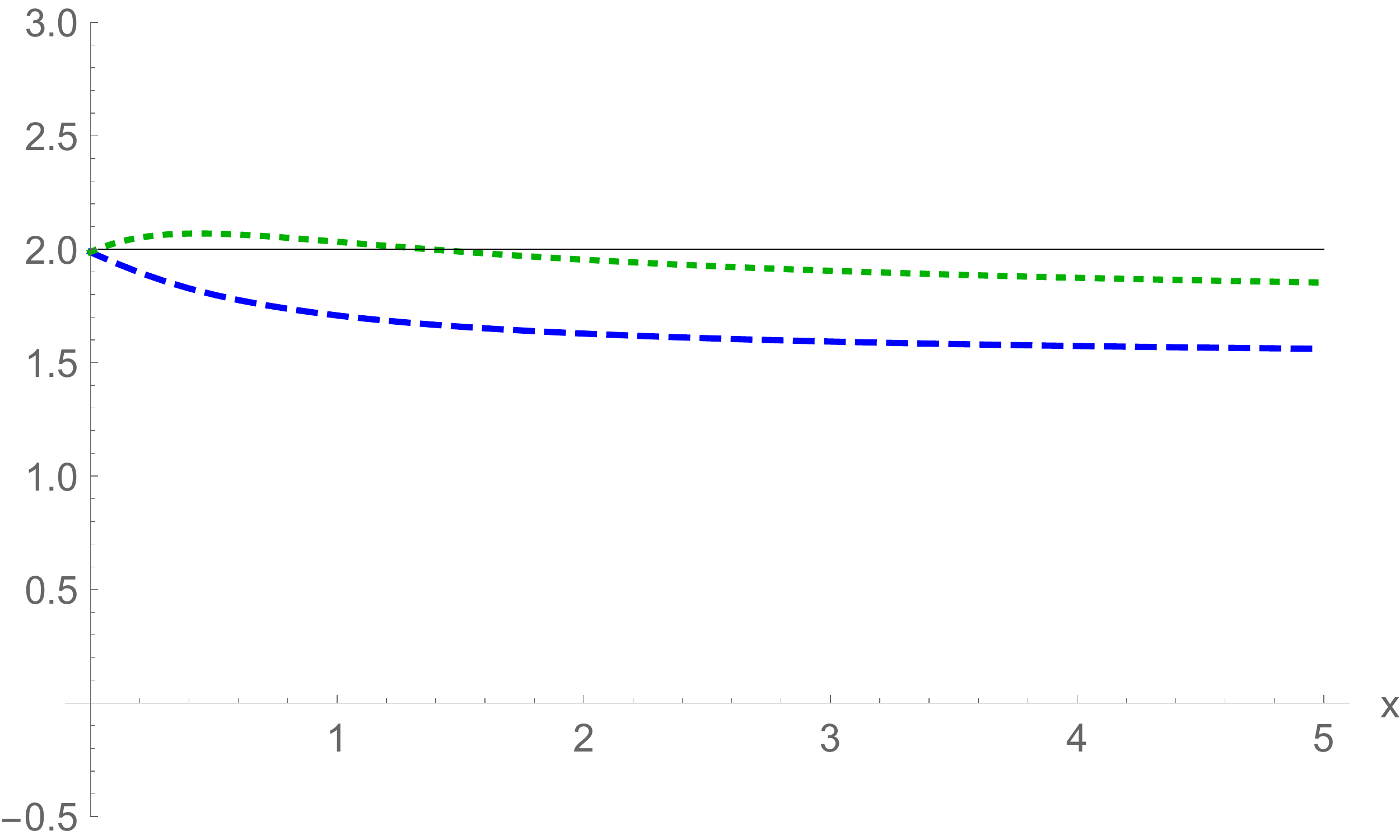}
\caption{Comparison of the violation of the CGLMP inequality in the 
	state $\ket{\xi(k,k^\pi)}$ (blue, dashed line) and in the state 
	$\ket{\psi(k,k^\pi)}$ (green, dotted line). The configuration of particles momenta
	and measurements directions is the following:
	$\vec{n}=(0,0,1)$, 
	$\vec{w}=(\cos\phi_w \sin\theta_w, \sin\phi_w \sin\theta_w, \cos\theta_w)$,
	$\vec{w}\in\{\vec{a},\vec{b},\vec{c},\vec{d}\}$ and
	$\theta_a= 3.141$, $\phi_a= 0$,
	$\theta_b= 0$, $\phi_b= 0$,
	$\theta_c= 0.836$, $\phi_c= 5.044$,
	$\theta_d= 2.754$, $\phi_d= 1.897$.} 
\label{fig:CGLMP_xi_psi-2}
\end{figure}

\begin{figure}
\includegraphics[width=0.95\columnwidth]{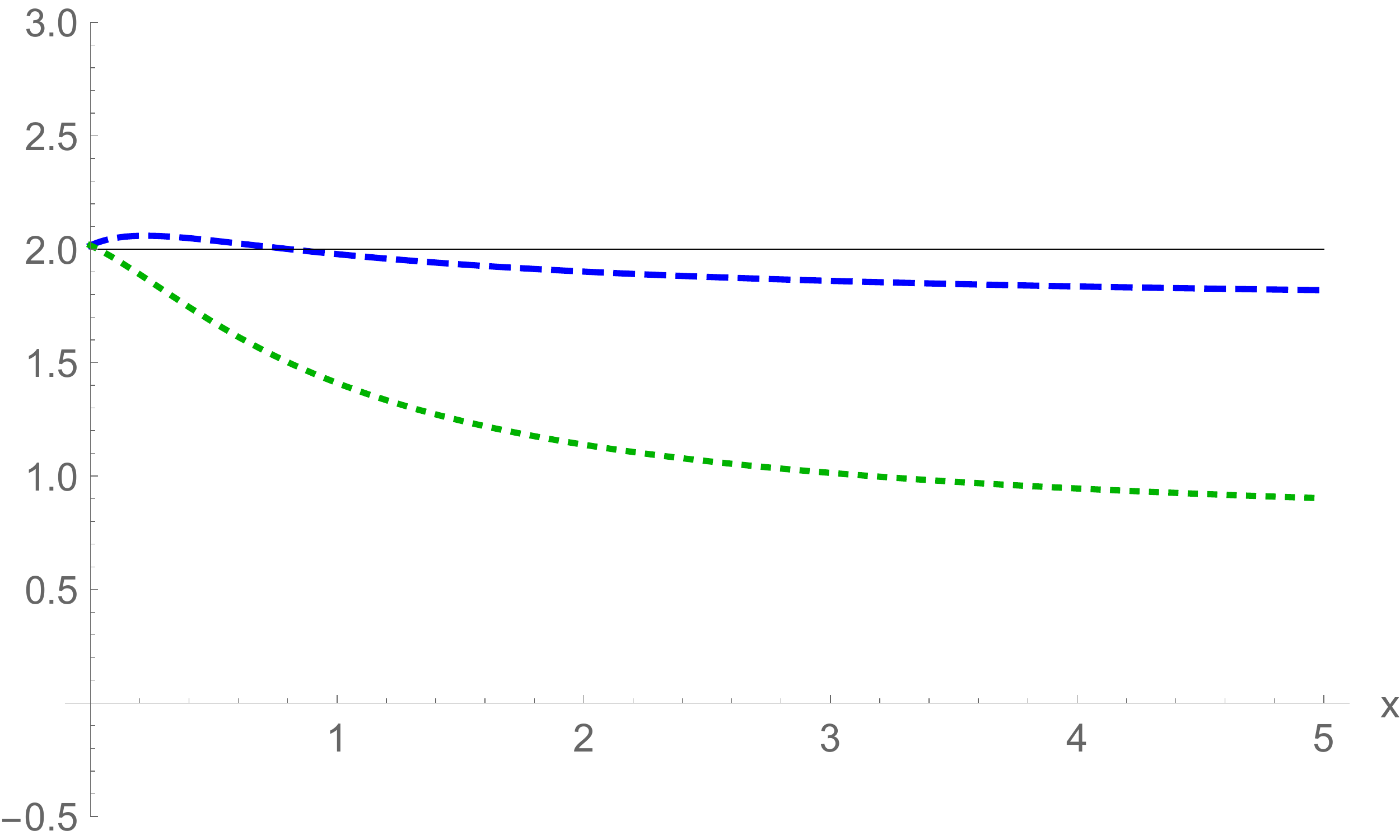}
\caption{Comparison of the violation of the CGLMP inequality in the 
	state $\ket{\xi(k,k^\pi)}$ (blue, dashed line) and in the state 
	$\ket{\psi(k,k^\pi)}$ (green, dotted line). The configuration of particles momenta
	and measurements directions is the following:
	$\vec{n}=(0,0,1)$, 
	$\vec{w}=(\cos\phi_w \sin\theta_w, \sin\phi_w \sin\theta_w, \cos\theta_w)$,
	$\vec{w}\in\{\vec{a},\vec{b},\vec{c},\vec{d}\}$ and
	$\theta_a= 2.532$, $\phi_a= 3.141$,
	$\theta_b= 1.213$, $\phi_b= 0$,
	$\theta_c= 2.378$, $\phi_c= 1.363$,
	$\theta_d= 0$, $\phi_d= 0$.} 
\label{fig:CGLMP_xi_psi-3}
\end{figure}

\paragraph*{CGLMP inequality.}
The CGLMP inequality is an optimal inequality for detecting quantum nonlocality
in a system of two qudits. For two qubits it reduces to the CHSH inequality.
We consider here spin one bosons therefore we present
the CGLMP inequality for two qutrits. We assume that Alice can perform 
two possible measurements $A_1$ or $A_2$, and Bob can perform measurements
$B_1$ or $B_2$. Each of these measurements can have three outcomes: 0,1,2.
Denoting by $P(A_i=B_j+k)$ the probability that the outcomes $A_i$ and $B_j$ differ
by $k$ modulo 3, 
i.e., $P(A_i=B_j+k) = \sum_{l=0}^{l=2} P(A_i=l,B_j=l+k \mod 3)$,
and defining
\begin{multline}
\mathcal{I}_3 = \big[
P(A_1=B_1) + P(B_1=A_2+1)\\
+ P(A_2=B_2) +P(B_2=A_1)
\big]\\
-\big[
P(A_1=B_1-1) + P(B_1=A_2)\\
+P(A_2=B_2-1) + P(B_2=A_1-1)
\big],
\end{multline}
the CGLMP inequality can be written in the form
\begin{equation}
\mathcal{I}_3 \le 2.
\end{equation}
Identifying spin projection values $-1,0,1$ with outcomes $0,1,2$ in the following way
\begin{equation}
-1 \leftrightarrow 0, \quad
0 \leftrightarrow 1, \quad
1 \leftrightarrow 2,
\end{equation}
and measurements $A_1$, $B_1$, $A_2$, $B_2$ with spin projections on 
$\vec{a}$, $\vec{b}$, $\vec{c}$, $\vec{d}$, respectively, the $\mathcal{I}_3$
takes the form
\begin{multline}
\mathcal{I}_3 = 
C(\vec{a},\vec{b}) + C(\vec{c},\vec{d}) + C(\vec{a},\vec{d}) - C(\vec{c},\vec{b})\\
+ P_{+-}(\vec{a},\vec{b}) + P_{+-}(\vec{c},\vec{d}) + P_{-+}(\vec{a},\vec{d})
- P_{+-}(\vec{c},\vec{b})\\
+ P_{00}(\vec{a},\vec{b}) + P_{00}(\vec{c},\vec{d}) + P_{00}(\vec{a},\vec{d})
- P_{00}(\vec{c},\vec{b}) \\
-\big[
P_{0-}(\vec{a},\vec{b}) + P_{0-}(\vec{c},\vec{d}) + P_{-0}(\vec{a},\vec{d})
- P_{0-}(\vec{c},\vec{b})\\
+ P_{+0}(\vec{a},\vec{b}) + P_{+0}(\vec{c},\vec{d}) + P_{0+}(\vec{a},\vec{d})
- P_{+0}(\vec{c},\vec{b})
\big].
\label{I3_bosons}
\end{multline}
The probabilities and correlation functions in Eq.~(\ref{I3_bosons}) are given
in Eqs.~(\ref{probabilities-center-of-mass}) and (\ref{correl_function_general}).
We have found that bosons can violate 
the CGLMP inequality either in the state $\ket{\psi(k,k^\pi)}$ or
in the state $\ket{\xi(k,k^\pi)}$.
In Figs.~\ref{fig:CGLMP_xi_psi-1}, \ref{fig:CGLMP_xi_psi-2}, \ref{fig:CGLMP_xi_psi-3}
we have compared 
the violation of the CGLMP inequality for the states $\ket{\psi(k,k^\pi)}$ 
and $\ket{\xi(k,k^\pi)}$ in different configurations.
In all these figures the point $x=0$ corresponds to the nonrelativistic case.

\section{Conclusions}

The recent paper \cite{Barr2021} suggested that it might be possible to experimentally
test the violation of Bell inequalities with a pair of $W^+ W^-$ bosons.
Motivated by this paper and by our previous theoretical works
\cite{CRW_2008_vector_bosons,Caban_2008_bosons_helicity}
we have discussed the violation of the Mermin and CGLMP inequalities in a system
of two relativistic vector bosons in a scalar state. 
We have derived formulas for probabilities in the 
EPR-type experiment in the general scalar state (\ref{scalar-state-general}),
assuming that Alice and Bob measure spin projections on given directions.
These probabilities depend on spin projection direction and bosons momenta.
We have considered in detail the situation when Alice and Bob are at rest
in the center of mass frame of the boson pair. In this case we have explicitly
calculated probabilities in two states of particular interest: the simplest nonseparable
state $\ket{\psi}$ (\ref{scalar-state-psi}) and the state which 
in the massless limit converges to
the scalar two-photon state $\ket{\xi}$ (\ref{scalar-state-xi}). 
We have shown that both the Mermin and CGLMP inequalities can be violated
in both states $\ket{\psi}$ and $\ket{\xi}$ and that the degree of violation depends 
on bosons momenta.

Large violations of the CGLMP inequality are predicted even for boson pairs at rest. 
For small boosts of the vector bosons, both the Mermin, and more particularly 
the CGLMP inequalities show sufficiently large violations that they might well be 
measurable in practical experiments. For the CGLMP inequality, our analytical results, 
here calculated in the narrow-width approximation, and with observers at rest in the centre-of-mass-frame of the boson pair, support the conclusion found using numerical
results in simulations of $H\to W W^*$
decays in Ref.~\cite{Barr2021} under different assumptions.

Potential applications of these results extend beyond the case of Higgs boson decays. 
Other example applications might include relativistic hadronic, nuclear, atomic 
or molecular systems.

\begin{acknowledgments}
AJB is grateful to Christopher Hays and to Alexander Karlberg for helpful discussions. 
AJB is supported through STFC grants ST/R002444/1
and ST/S000933/1, and by Merton College, Oxford.
PC and JR are supported by University of Lodz under the IDUB project.
\end{acknowledgments}

\appendix

\section{Massive representations of the Poincar\'e group for $s=1$}
\label{sec:app-Poincare}

For self-consistency we recall here basic facts related to
massive spin 1 representations of the Poincar\'e group.
For more details see, e.g, \cite{BR1986-Groups-book}.
Let $\mathcal{H}$ be the carrier space of the
irreducible massive representation of the Poincar\'e group for $s=1$. 
$\mathcal{H}$ is spanned by the four-momentum operator eigenvectors
$\ket{k,\sigma}$
\begin{equation}
\label{four-momentum}
\hat{P}^{\mu}\ket{k,\sigma}=k^{\mu}\ket{k,\sigma},
\end{equation}
$k^2=m^2$, $m$ is the mass of the particle, and $\sigma$
its spin component along $z$ axis, $\sigma=-1,0,1$. 
We use the following Lorentz-covariant normalization
\begin{equation}
	\label{normalization_base}
	\bracket{k,\sigma}{k',\sigma'}=
	2k^0\delta^3(\vec{k}-\vec{k}^{\prime})\delta_{\sigma\sigma'}.
\end{equation}
The vectors $\ket{k,\sigma}$ can be generated from the standard vector
$\ket{\tilde{k},\sigma}$, where $\tilde{k}=m(1,0,0,0)$ is the
four-momentum of the particle in its rest frame. We have
$\ket{k,\sigma}=U(L_k)\ket{\tilde{k},\sigma}$, where Lorentz boost
$L_k$ is defined by relations $k=L_k\tilde{k}$,
$L_{\tilde{k}}=\id$.

By means of Wigner procedure we get
\begin{equation}
\label{U_Lambda_base}
U(\Lambda)\ket{k,\sigma}=
\mathcal{D}_{\lambda\sigma}(R(\Lambda,k))\ket{\Lambda	k,\lambda},
\end{equation}
where the Wigner rotation $R(\Lambda,k)$ is defined as
$R(\Lambda,k)=L_{\Lambda k}^{-1}\Lambda L_k$, and
for $s=1$ the representation $\mathcal{D}(R)$ is unitary equivalent
to $R$ by
\begin{equation}
	\label{equivalent}
	\mathcal{D}(R)=VRV^{\dag},\qquad V^\dagger V = \id,
\end{equation}
and the explicit form of the matrix $V$ is the following:
\begin{equation}
	\label{matrix_V}
	V=\frac{1}{\sqrt{2}}
	\begin{pmatrix}
		-1 & i & 0 \\
		0 & 0 & \sqrt{2} \\
		1 & i & 0 \\
	\end{pmatrix}.
\end{equation}

\section{Boson field }
\label{sec:app-field}

In order to describe two types of vector bosons, particle and antiparticle 
(e.g. $W^+$ and $W^-$), we consider
the free field operator $\Hat{\phi}^\mu(x)$ with the following momentum expansion:
\begin{multline}
\Hat{\phi}^\mu(x) = (2\pi)^{(-3/2)} \int \tfrac{d^3 \vec{k}}{2\omega_k}
\big[
e^{ikx} e^{\mu}_{\sigma}(k) a^{\dagger}_{\sigma}(k)\\
+ e^{-ikx} e^{*\mu}_{\sigma}(k) b_{\sigma}(k)
\big],
\end{multline}
where $\omega_k=\sqrt{\vec{k}^2+m^2}$ and $m$ is a mass of a particle 
(and antiparticle). $a^{\dagger}_{\sigma}(k)$, $a_\sigma(k)$ and
$b^{\dagger}_{\sigma}(k)$, $b_\sigma(k)$ are creation and annihilation operators
of a particle and antiparticle, respectively. 
$a^{\dagger}_{\sigma}(k)$ ($b^{\dagger}_{\sigma}(k)$) creates
particle (antiparticle) with four-momentum $k$ and spin component along $x$ axis
equal to $\sigma$. They fulfill the standard canonical commutation relations
\begin{equation}
[a_{\sigma}(k),a^{\dagger}_{\sigma^\prime}(k^\prime)] = 
[b_{\sigma}(k),b^{\dagger}_{\sigma^\prime}(k^\prime)] = 
2 k^0 \delta(\vec{k}-\vec{k}^\prime) \delta_{\sigma\sigma'}
\end{equation}
and all the other commutators vanish.
The Klein-Gordon equation and Lorentz transversality condition imply
\begin{equation}
k^2=m^2, \quad k_\mu e^{\mu}_{\sigma}(k)=0.
\label{e-transversality}
\end{equation}

The one-particle and one-antiparticle states
\begin{equation}
\ket{k,\lambda}_a = a^{\dagger}_{\lambda}(k) \ket{0}, \quad
\ket{p,\sigma}_b = b^{\dagger}_{\sigma}(p) \ket{0},
\end{equation}
where $\ket{0}$ is a Lorentz-invariant vacuum, $\bracket{0}{0}=1$, 
should transform like
a basis states of a carrier space of the irreducible, massive representation
of the Poincar\'e group for $s=1$ considered in Appendix \ref{sec:app-Poincare}.
This condition allows us to determine the explicit form of amplitudes 
$e^{\mu}_{\sigma}(k)$. The derivation is the same as 
in \cite{CRW_2008_vector_bosons}, therefore we give here only the results:
\begin{equation}
e(k) = [e^{\mu}_{\sigma}(k)] = 
\begin{pmatrix}
\tfrac{\vec{k}^T}{m}\\
\id + \tfrac{\vec{k}\otimes \vec{k}^T}{m(m+k^0)}
\end{pmatrix}
V^T, 
\label{amplitude-e-explicit}
\end{equation}
where $V$ is given in Eq.~(\ref{matrix_V}). Moreover, 
one can show that $e^{\mu}_{\sigma}(k)$ fulfills the following 
relations:
\begin{align}
& e^*(k) = e(k) V V^T,\\
& e^{*\mu}_{\sigma}(k) e^{\nu}_{\sigma}(k) = - \eta^{\mu\nu} 
+ \tfrac{k^\mu k^\nu}{m^2},\\
& e^{*\mu}_{\sigma}(k) e_{\mu\lambda}(k) = - \delta_{\sigma\lambda},\\
& e^{\mu}_{\sigma}(k) e_{\mu\lambda}(k) = - (VV^T)_{\sigma\lambda}.
\end{align}

\section{Boson states transforming covariantly under Lorentz transformations}
\label{sec:app-states}

In \cite{CRW_2008_vector_bosons} it was shown that 
with the help of amplitudes $e^{\mu}_{\sigma}(k)$ one can construct states
transforming in the explicitly covariant manner under Lorentz transformations.
In our case covariant particle/anti-particle states have the form
\begin{equation}
\ket{(\mu,k)}_{a/b} = e^{\mu}_{\sigma}(k) \ket{k,\sigma}_{a/b},
\label{covariant_states}
\end{equation}
and under Lorentz transformations transform according to
\begin{equation}
U(\Lambda) \ket{(\mu,k)}_{a/b} = 
(\Lambda^{-1})^{\mu}_{\nu} \ket{(\nu,\Lambda k)}_{a/b}.
\end{equation}
Two-particle state describing boson with four-momentum $k$ and 
spin projection $\lambda$ and antiboson with four-momentum $p$ and spin
projection $\sigma$ has the form
\begin{equation}
\ket{(k,\lambda)_a;(p,\sigma)_b} = 
a^{\dagger}_{\sigma}(k) b^{\dagger}_{\sigma}(p) \ket{0},
\label{state-two-boson}
\end{equation}
and consequently a boson--antiboson covariant state reads
\begin{equation}
e^{\mu}_{\lambda}(k) e^{\nu}_{\sigma}(p) 
\ket{(k,\lambda)_a;(p,\sigma)_b}.
\label{covariant-state-two-boson}
\end{equation}

\section{Localization inside detectors}
\label{sec:localization}

Eigenvectors of the Newton--Wigner operator have the following
form\cite{CRRSW2014_correl-localization}
\begin{equation}
		\ket{\vec{x},\sigma} = \frac{1}{(2\pi)^{3/2}} 
		\int \frac{d^3 \vec{p}}{2p^0} \sqrt{2p^0} 
		e^{- i \vec{p}\cdot\vec{x}} \ket{p,\sigma},
		\label{NW-eigenvector}
\end{equation}
where $p^0=\sqrt{\vec{p}^2+m^2}$. 
The problem of localization of 
a particle inside a detector during spin measurement 
in the context of relativistic quantum correlations of a fermion pair
was also discussed in \cite{CRRSW2014_correl-localization}.
Here we briefly present that paper's conclusions, adapting them 
to a system of vector bosons.
The central element in such a discussion is the projector
on a region $D$ in the coordinate space
\begin{align}
		\Hat{\Pi}_D & = \int_D d^3 \vec{x} \sum_{\sigma=-1,0,1} 
		\ket{\vec{x},\sigma}\bra{\vec{x},\sigma}\\
		& = \iint\limits_{{{\mathbb{R}}^3\times{\mathbb{R}}^3}}
		\frac{d^3 \vec{p}}{\sqrt{2p^0}}
		\frac{d^3 \vec{p}^\prime}{\sqrt{2{p^{\prime}}^0}}
		\Delta_D(\vec{p}^\prime-\vec{p})
		\sum_{\sigma} \ket{p^\prime,\sigma}\bra{p,\sigma},
\end{align}
where $\ket{\vec{x},\sigma}$ is defined in (\ref{NW-eigenvector}) and
\begin{align}
	\Delta_D(\vec{p}^\prime-\vec{p}) & = \frac{1}{(2\pi)^3}
	\int_D d^3 \vec{x} e^{-i (\vec{p}^\prime - \vec{p})\cdot \vec{x}}\\
	& = \frac{1}{(2\pi \hbar)^3}
	\int_D d^3 \vec{x} e^{-i \frac{(\vec{p}^\prime - \vec{p})\cdot \vec{x}}{m c \lambda}}.
\end{align}
To facilitate further discussion, in the second line we have explicitly expressed 
$\Delta_D$ in the standard units,
$c$ is the velocity of light and $\lambda=\frac{\lambda_C}{2\pi}=
\frac{\hbar}{mc}$ is the particle Compton wavelength divided by $2\pi$.
To evaluate quantitatively the influence of localization let us choose $D$ as the 
cube located in the center of the coordinate frame with edges of the length $l$ 
parallel to coordinate axis. 
In this case $\Delta_\textsf{Cube}$ can be calculated and we obtain
\begin{equation}
		\Delta_\textsf{Cube}(\vec{p}^\prime-\vec{p}) = 
		\Big(\frac{\lambda}{\hbar}\Big)^3 
		\prod_{j=1}^{3} 
		\Big[
		\frac{1}{\pi} \frac{l}{2\lambda} \mathrm{sinc}\,
		\Big(
		\frac{p_{j}^{\prime} - p_j}{mc} \frac{l}{2\lambda}
		\Big)
		\Big].
\end{equation}
Using the formula
\begin{equation}
		\lim\limits_{\tau\to\infty} \frac{\tau}{\pi} \mathrm{sinc}\,(\tau x)
		=\delta(x)
\end{equation}
we observe that for $l\to\infty$:
$\Delta_\textsf{Cube}(\vec{p}^\prime-\vec{p})\to
\delta^3(\vec{p}^\prime-\vec{p})$
and consequently
\begin{equation}
		\Hat{\Pi}_D \to \Hat{\Pi}_{{\mathbb{R}}^3}	=I.
\end{equation}
Therefore, there is no problem with localization of a quantum particle 
provided that its Compton wavelength is sufficiently small in comparison with 
a detecting element size $l$, i.e., when the condition $\lambda\ll l$ holds.
	
For muons and electrons at rest the corresponding Compton
wavelengths take the following approximate values:
$\lambda_\mu = 1.87\times10^{-15}$~m and
$\lambda_e = 3.86\times10^{-13}$~m.
Consequently, assuming a realistic size of particle detector pixels
$l = 10^{-6}$~m, 
the scaling factors $\tau=l/(2\lambda)$ take the values
$\tau_\mu = 0.27\times10^{9}$ and
$\tau_e = 0.13\times10^{7}$,
respectively.
Moreover, for relativistic particles we should use 
$m = \gamma m_\textsf{rest}$, where $\gamma$ is the particle Lorentz factor, 
so $\lambda = \gamma^{-1} \lambda_\textsf{rest}$ and in effect 
the Compton wavelength decreases so the scaling factor increases.
Thus, for real, massive particles $\Delta_\textsf{Cube}(\vec{p}^\prime-\vec{p})$
is very close to $\delta^3(\vec{p}^\prime-\vec{p})$
and, indeed, there is no problem with localization of a particle inside the detector
in situation in which it is directly detected.


\end{document}